\documentclass[letterpaper, 10 pt, journal, twoside]{ieeetran}

\usepackage{amsmath,amsfonts,graphicx,multirow,enumitem,tabularx}

\author{Mingjie Bi$^{1}$, Juan-Alberto Estrada-Garcia$^{2}$, Dawn M. Tilbury$^{1,3}$, Siqian Shen$^{2}$ and Kira Barton$^{1,3}$
\thanks{Manuscript received: October, 27, 2023; Revised February, 8, 2024; Accepted April, 5, 2024.}
\thanks{This paper was recommended for publication by Editor Chao-Bo Yan Name upon evaluation of the Associate Editor and Reviewers' comments.
This work was supported by The United States National Science Foundation (NSF) grant \#CMMI-2034974.} 
\thanks{$^{1}$Mingjie Bi was with the Robotics Department, 
        University of Michigan, Ann Arbor, MI 48109, USA,
        {\tt\footnotesize mingjieb@umich.edu}}%
\thanks{$^{2}$Juan-Alberto Estrada-Garcia and Siqian Shen are with the Department of Industrial and Operations Engineering, 
        University of Michigan, Ann Arbor, MI 48109, USA,
        {\tt\footnotesize \{juanest, siqian\}@umich.edu}}%
\thanks{$^{1,3}$Dawn M. Tilbury and Kira Barton are with the Robotics Department and the Department of Mechanical Engineering, 
        University of Michigan, Ann Arbor, MI 48109, USA,
        {\tt\footnotesize \{tilbury, bartonkl\}@umich.edu}}%
\thanks{Digital Object Identifier (DOI): see top of this page.}
}

\title{Heterogeneous Risk Management Using a Multi-agent Framework for Supply Chain Disruption Response}

\allowdisplaybreaks
\begin{document}
\maketitle
\begin{abstract}

In the highly complex and stochastic global, supply chain environments, local enterprise agents seek distributed and dynamic strategies for agile responses to disruptions.
Existing literature explores both centralized and distributed approaches, while most work neglects temporal dynamics and the heterogeneity of the risk management of individual agents.
To address this gap, this paper presents a heterogeneous risk management mechanism to incorporate uncertainties and risk attitudes into agent communication and decision-making strategy.
Hence, this approach empowers enterprises to handle disruptions in stochastic environments in a distributed way, and in particular in the context of multi-agent control and management. 
Through a simulated case study, we showcase the feasibility and effectiveness of the proposed approach under stochastic settings and how the decision of disruption responses changes when agents hold various risk attitudes.

\end{abstract}


\section{introduction}
\label{sec:introduction}
\IEEEPARstart{T}{he} growing complexity of the global supply chain brings uncertainties to all the stages from supply and manufacturing to storage and delivery~\cite{simangunsong2012supply}.
These uncertainties frequently disrupt the supply chain network (SCN) and result in, e.g., supply delays and shifting demands~\cite{xu2020disruption}.
Existing literature mostly focuses on proactive methods for supply chain risk mitigation by estimating potential risks in advance to enhance supply chain robustness~\cite{ivanov2017literature}. 
Sometimes, responses need to be conducted in real-time and adaptive to disruptive scenarios, but few studies incorporate heterogeneous risk management into this re-planning process~\cite{baryannis2019supply}.

In the supply chain domain, most research focuses on centralized risk management to provide optimal solutions based on specific objectives (e.g., minimizing total cost)~\cite{ivanov2017literature}.
These approaches consider holistic risk at the supply chain level, and therefore for disruption response, centralized models require full supply-chain information. 
As the complexity and scale of supply chains increase, it becomes more difficult to adopt centralized approaches to effectively respond to disruption and manage risk~\cite{baryannis2019supply, kovalenko2022toward}.

Modern supply chains are heterogeneous, where agents in the SCN play different roles and may possess diverse objectives~\cite{xu2021will}.
Therefore, for risk management, it is important to distinguish the agents' risk attitudes if necessary, which may also change dynamically as the supply chain environment evolves.
To enable the consideration of heterogeneity and improve the agility of supply chain risk management, researchers have adopted distributed approaches, where multiple entities in the system make decisions via local communication and collaboration~\cite{xu2021will, toorajipour2021artificial}.

Multi-agent control is a commonly-used distributed method that enables intelligent decision-making in manufacturing systems and supply chains~\cite{bi2021dynamic, bi2023dynamic, bi2023distributedthesis}.
Each autonomous agent in an SCN either represents a physical entity (e.g., a supplier) or is responsible for a function (e.g., demand forecasting). %
Agents communicate, make local decisions, and collaboratively solve supply chain problems~\cite{toorajipour2021artificial, rahimi2018auction}.
Agents consider different risk-based models to solve their local problems,
depending on available information and risk attitudes. 
Such flexibility allows heterogeneous risk evaluation which is complex in large-scale centralized formulations.

Most existing agent-based disruption response strategies are case-based and rule-based decision-making~\cite{blos2018framework,giannakis2011multi, ghadimi2019intelligent}, and thus require prior knowledge of disruptions and reactive actions.
They focus on identifying risk mitigation actions from the system level without considering individual agent risk management, or uncertainties in response actions.
Some studies derive individual agent risk models but do not consider the disruption response problem~\cite{wu2021impact, kwon2007mace}.
In~\cite{zhang2022pareto,zhou2018cooperative, zang2022coordinating, choi2019optimal}, agents are assumed to have identical risk attitudes and the work in~\cite{heidary2019risk} allows different agent risk attitudes but only for agents who are customers.
However, these approaches do not make full use of the heterogeneity of multi-agent systems to conduct heterogeneous and dynamic risk management. 

The main contributions of this paper include: (1) the development of a heterogeneous and dynamic risk management mechanism for supply-chain agents, (2) the incorporation of risk-aware stochastic optimization for agent decision-making in response to supply-chain disruptions, and (3) an evaluation of the SCN performance with various agent risk attitudes through a simulation-based case study considering out-of-sample disruption scenarios.

The rest of the paper is organized as follows. 
In Section~\ref{sec:formulation}, we present an overview of the multi-agent SCN and problem formulation.
In Section~\ref{sec:heterogeneous}, we introduce a heterogeneous risk management mechanism.
In Section~\ref{sec:disruptionresponse}, we describe the agent communication and decision-making considering risk.
In Section~\ref{sec:casestudy}, we present numerical results via a case study and provide concluding remarks in Section~\ref{sec:conclusion}.

\section{Problem Description and formulation}
\label{sec:formulation}

\subsection{Multi-agent SCN}
\label{sec:agentnetwork}

We denote an SCN as $G(V,E)$, where $V$ is the set of vertices, representing supply chain entities, such as suppliers, customers, etc., and $E$ is the set of edges, representing product/material flows between the entities. 
All the vertices and edges are associated with additional information (e.g., lead time, production, cost, and capacity) to describe their characteristics. 
Both vertices and edges in the SCN have intelligence and thus are defined as agents.
We denote the corresponding agent network by $G^a(A,L)$, where $A=V\cup E$ is the set of agents, including all the vertices in $V$ and edges in $E$.
The agents in $A$ are connected by a set $L$ of communication links. 
In the multi-agent SCN, agents have self-awareness about their own attributes, can communicate to share information, and make decisions based on their knowledge and goals.
We consider the following agent types: customer, distributor, original equipment manufacturer (OEM), tier supplier, and transporter.

\subsection{Re-planning problem formulation}
\label{sec:prblemformulation}

We focus on the following problem:
given an SCN, individual agents, existing product flows, and a stochastic disruptive event at an agent, how can we model and incorporate agents' risk attitudes into the decision-making to improve the resiliency of the disruption response?
We make the following assumptions to specify our scope:
\begin{enumerate}[label={A.\arabic*}]
    \item The given supply chain starts from an original plan that is pre-determined and planned.\label{a1}
    \item An unexpected disruption increases an agent's lead time and delays delivery, and can be detected immediately.  \label{a2}
    \item The uncertain parameters are production and lead time with known probability distributions inferred from historical data. \label{a3}
\end{enumerate}

\ref{a1} indicates that the supply chain follows an original optimal flow plan before the disruption occurs.
We describe the plan as all of the product flows ($y_{ijk}$) and arrival times ($v_{ijk}$) from agent $a_i$ to $a_j$ for product $k$ in the network.
\ref{a2} guarantees that a disruption will be identified by the agent when it occurs and also designates how the network will be impacted by the disruption. 
\ref{a3} enables the quantification of risks that are incorporated into agent decision-making. 

The disruption will trigger agents to re-organize the flow plan to minimize the production and flow costs, as well as the penalties.
Agents penalize the risk of demand dissatisfaction regarding product amount and delivery time due to the disruption in lead time and delays in the delivery~\cite{estradagarcia2023multiobjective}:
\begin{subequations}
\label{eq:centralized}
\begin{align}
\min_{\hat{y}, \hat{v}}\ & \mathcal{H}_p(\hat{y}, \hat{v}) + \mathcal{H}_t(\hat{y},\hat{v})\label{eq:centrobj}\\
\text{s.t.}\quad 
& \text{Agent constraints} \label{eqc:constr1}\\
& \text{Network constraints} \label{eqc:constr2}
\end{align}
\end{subequations}
where $\hat{y}$ and $\hat{v}$ represent the new flows and arrival times; $\mathcal{H}_p(\hat{y}, \hat{v})$ and $\mathcal{H}_t(\hat{y},\hat{v})$ compute the unmet demand and delivery lateness for all customers.
Instead of resolving the centralized model, we apply an agent-based distributed approach that is proposed in~\cite{bi2022model} to provide a new flow plan.
We revise the agent optimization by incorporating the uncertainties of production capacity and lead time.
Agents have different ways of handling uncertainties in the constraints and calculating the objective based on their own risk attitudes.

\section{Heterogeneous risk management}
\label{sec:heterogeneous}

\begin{figure}[tb]
\centerline{\includegraphics[width=0.75\columnwidth]{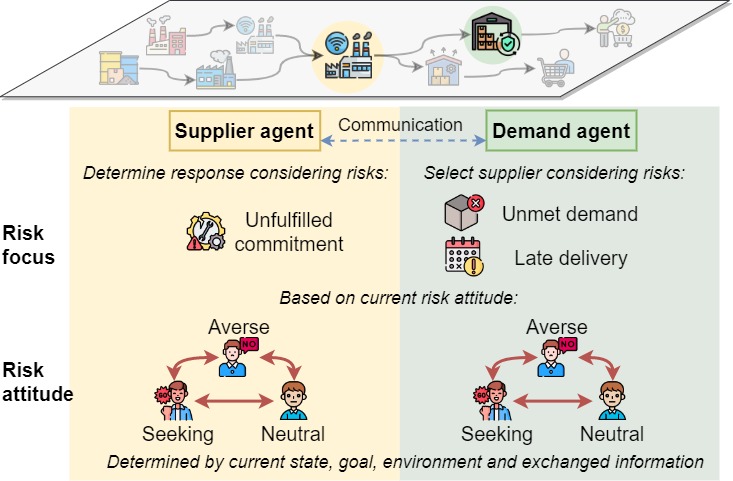}}
\caption{Risk management mechanism with heterogeneous risk focuses and risk attitudes for the roles of supplier agents and demand agents.}
\label{fig:heter}
\end{figure}

We introduce a heterogeneous risk management mechanism for SCNs to guide the communication and decision-making for a disruption event.
Each individual agent considers their own risks and applies a different risk attitude depending on their role in the SCN and their current status.
We define two types of risk heterogeneity: agent risk focus and agent risk attitudes, as shown in Fig.~\ref{fig:heter}.

\subsection{Heterogeneity of agent risk focus}
\label{sec:uncertaintyrisk}
Focusing on the re-planning problem for disruption response, we define two roles, supplier agents and demand agents.
The supplier agents receive demand requests and provide products if they are selected.
The demand agents need a certain amount of products at a given time to guarantee their scheduled production plans.
Every agent in the SCN can be both a supplier agent and a demand agent in different scenarios.
Therefore, agents consider different risks when they play different roles, which results in a dynamic and typically heterogeneous risk focus across the network.

\subsubsection{Risks for supplier agents}
when supplier agent $a_z$ receives demand requests from multiple demand agents, it makes decisions on how it responds to these requests by evaluating their capabilities.
Agents seek maximal income and rewards while considering the risk of failing to fulfill a commitment and the penalty associated with such failure.
Each supplier agent maximizes the following function (additional details are included in Section~\ref{sec:requestresponse}):
\begin{equation}
    \mathcal{J}_s = \sum_{a_j\in \mathcal{A}_{dm}, k\in K}r_{zjk}\bar{y}_{zjk} + w_s^e\sum_{a_j\in \mathcal{A}_{dm}}g_j\eta_j-w_s^rR_s, 
    \label{eqn:supobj}
\end{equation}
where $r_{zjk}$ represents the income per unit that the supplier can earn from demand agent $a_j$ for product $k$ and $\bar{y}_{zjk}$ is the decision variable representing the quantity that supplier $a_z$ plans to commit to demand agent $a_j$ for product $k$.
Parameter $g_j$ indicates the rewards that supplier agents can gain from demand agent $a_j$ if the product fulfillment is satisfied ($\eta_j=1$).
The rewards could include a bonus or future contract if the supplier agents can satisfy all the demands.
Variable $R_s$ evaluates the risk of not fulfilling the response due to production and lead time uncertainties.
The weights $w_s^e$ and $w_s^r$ control the importance of rewards and risks for each agent.
Supplier agents respond to disruptions by evaluating the risk tolerance of unfulfilled commitments, managing the trade-off between risk and demand fulfillment.
The details of the optimization model are in Section~\ref{sec:requestresponse}.

\subsubsection{Risks for demand agents}
Demand agents set their objective \eqref{eqn:demobj} for supplier selection considering the response they get from supplier agents with respect to product availability and unmet demand and lateness penalties.
\begin{equation}
\vspace{-5pt}
    \mathcal{J}_d = C_d + w_d^t\sum_{k\in K}\Delta_{jk}^t + w_d^p\sum_{k\in K}\Delta_{jk}^p,
    \label{eqn:demobj}
\end{equation}
where $C_d$ represents the cost of obtaining products from supplier agents; $\Delta_{jk}^t$ represents the delay times, and $\Delta_{jk}^p$ represents the amount of unmet demand of product $k$; the weights $w_d^t$ and $w_d^p$ are used to weigh the importance of unmet demand versus delay time for the specific demand agent. 
Different demand agents can have different weights or alternative objectives and risks to consider.
Depending on the uncertainty associated with a specific supplier's production capacity or lead time, demand agents will evaluate the associated risks and make a selection decision depending on their perceived risk attitude. 
The details of the optimization model are in Section~\ref{sec:supplierselection}.

\subsection{Heterogeneity of agent risk attitude}
\label{sec:agentheterogeneity}

In addition to the risk focus, agents can have different risk attitudes, which represent how agents balance risks and their original performance objectives (e.g., cost, revenue) depending on their current status.
We consider states $X_r=\{averse, neutral\}$ to represent the different risk attitudes.
Specifically, an \textit{averse} state indicates that agents try to make conservative decisions, i.e., avoid deviations between their behaviors and decisions.
For agents in a \textit{neutral} state, their decision-making aims to balance their objectives and risk assessment and avoid both conservative and risky decisions.
Therefore, risk attitudes correspond to how agents measure the consequences of uncertainty. These  attitudes may change as agents communicate and make decisions.
For example, an agent could be risk-neutral when it supplies products but risk-averse when it demands products.
Also, a supplier agent could be risk-neutral when it has an optimistic estimation of its production but risk-averse if not.
Therefore, our approach allows heterogeneous and dynamically-changing risk focus, attitudes, and tolerance for each agent. 

\section{Disruption response with risk assessment}
\label{sec:disruptionresponse}

\begin{figure}[tb]
\centerline{\includegraphics[width=\columnwidth]{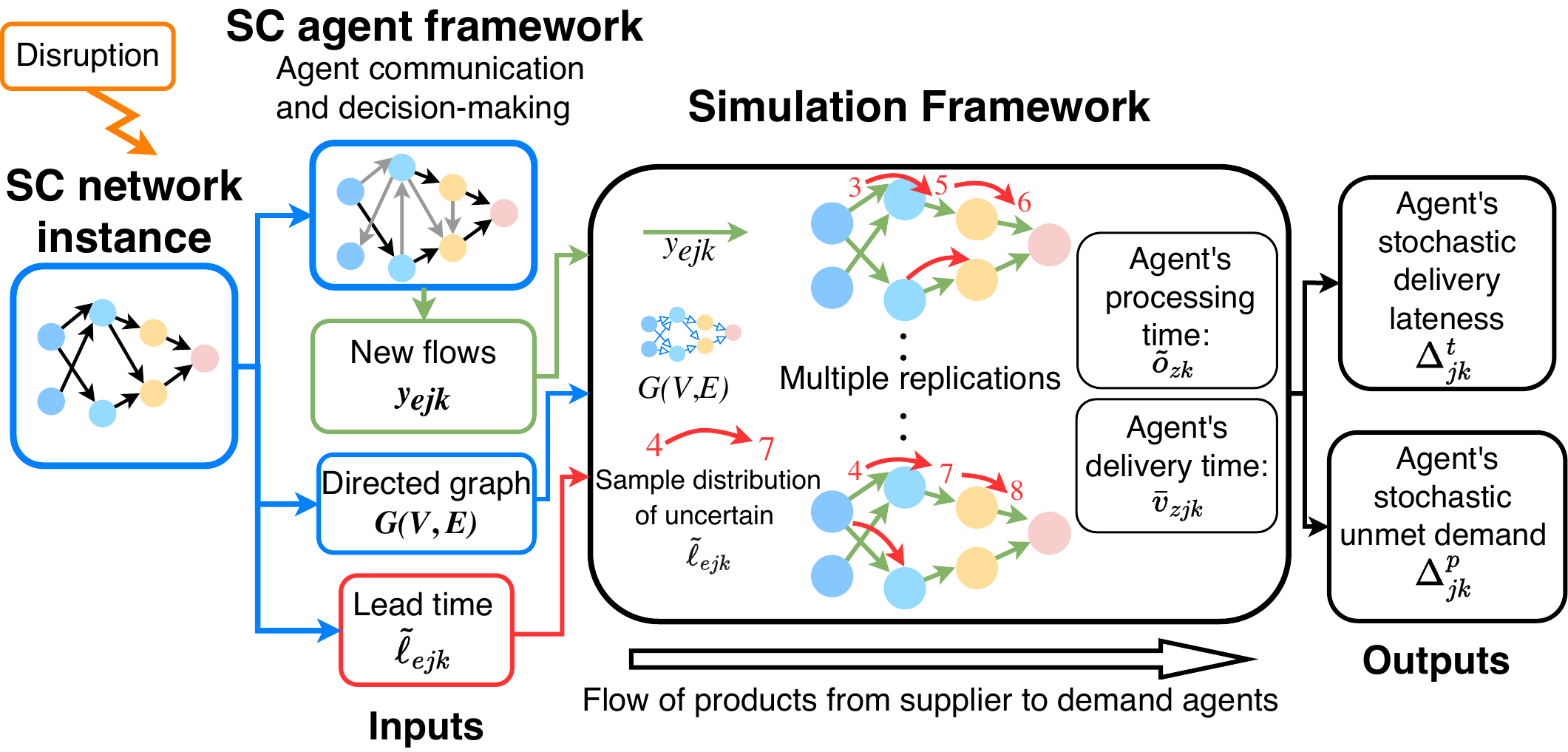}}
\caption{The agent-based decision optimization and simulation. The framework starts from input data collection to agent-based communication and decision-making, and then the solutions will be simulated under different disruption scenarios to obtain outcomes and evaluation results related to delivery lateness and demand shortage.}
\label{fig:framework}
\end{figure}

Our prior work introduces an agent-based framework describing the agent communication and decision-making schemes in response to SCN disruptions (see~\cite{bi2022model, bi2023distributed}).
In this paper, we focus on modeling agent decision making considering uncertainty and heterogeneous risk attitudes.
As shown in Fig.~\ref{fig:framework}, the agent-based framework serves as a distributed optimization model to provide a new flow plan for simulation-based evaluation.

\subsection{Disruption identification}
\label{sec:disruptionid}

We define $y_{ejk}$ as the amount of product $k$ that agent $a_e$ is scheduled to provide to agent $a_j$, and $v_{ejk}$ represents the arrival time associated with flow $y_{ejk}$.
Once the disruption occurs, the disrupted agent $a_e$ informs all the downstream agents $a_j$ about the new arrival time $v'_{ejk}$.

\subsection{Agent request and response}
\label{sec:requestresponse}

\subsubsection{Request}
From the original plan, a downstream agent $a_j$ is scheduled to receive product $k$ and use it to produce product $k'$ starting at time $o_{jk'}$.
Once agent $a_j$ receives the information about the new arrival time, it checks whether its production is affected by the lead time disruption.
If $v'_{ejk}>o_{jk'}$ (i.e., the product $k$ is late), then $a_j$ becomes a demand agent that seeks to obtain product $k$ from alternative supplier agents.
Otherwise, $a_j$ can accommodate the disruption and no re-planning decisions are needed.
We denote the set of all demand agents as $\mathcal{A}_{dm}$.
Each demand agent $a_j\in \mathcal{A}_{dm}$ sends a request for product $k$ to its upstream suppliers ($a_z\in Z_j(k)$) based on its environment knowledge.
The request includes $(d_{jk},t_{jk})$, where $d_{jk}=y_{ejk}$ denotes the demand amount for product $k$, which equals the flow from disrupted agent $a_e$, and $t_{jk}=o_{jk'}$ is the delivery deadline, which equals the planned production start time.
In the upper block in Fig.~\ref{fig:communication}, step 1 represents the request process.

\subsubsection{Response}

For each agent $a_z$, the response decisions include $(\bar{y}_z, \bar{v}_z)$, where $\bar{y}_z$ represents the number of products that $a_z$ is willing to provide and $\bar{v}_z$ represents the time at which it can deliver the products.
We allow agents to have production over their nominal production capacity, although these production commitments generally correspond to longer lead times.
Specifically, $\bar{y}_z=[\bar{y}_{zjk}^u, \bar{y}_{zjk}^o, \forall a_j\in \mathcal{A}_{dm}, k\in K]^{\mathsf{T}}$, where $\bar{y}_{zjk}^u$ and $\bar{y}_{zjk}^o$ represents the maximum units of product $k$ that $a_z$ can provide to $a_j$ within and over its production estimations, respectively. 
The arrival times for $\bar{y}_{zjk}^u$ and $\bar{y}_{zjk}^o$ are different: $\bar{v}_z=[\bar{v}_{zjk}^u, \bar{v}_{zjk}^o, \forall a_j\in \mathcal{A}_{dm}, k\in K]^{\mathsf{T}}$, where $\bar{v}_{zjk}^u$ and $\bar{v}_{zjk}^o$ represent the arrival times of $\bar{y}_{zjk}^u$ and $\bar{y}_{zjk}^o$ from $a_z$ to $a_j$, respectively.

Agent $a_z$ estimates the amount of product $k$ it can produce, denoted by $\tilde{p}_{zk}$, and the time it can start production, denoted by $\tilde{o}_{zk}$.
Both $\tilde{p}_{zk}$ and $\tilde{o}_{zk}$ are random variables with known distributions.
We formulate the optimization for the response decision-making as a stochastic programming problem, as shown in~\eqref{eq:response}.
This model maximizes the weighted sum of income and rewards, subtracting normalized risk values:

\begin{subequations}
\vspace{-10pt}
\label{eq:response}
\begin{align}
\max_{\bar{y}_z, \bar{v}_z}\hspace{0.1cm}
&\mathbb{E} \Bigg[\sum_{a_j\in \mathcal{A}_{dm}, k\in K}r_{zjk}(\bar{y}^u_{zjk}+\bar{y}^o_{zjk}) - w^r\bar{y}^o_{zjk} \nonumber\\
& \hspace{0.3cm}+ \sum_{a_j\in \mathcal{A}_{dm}}w^pg_j^p\prod_{k\in K}\eta^p_{zjk}+w^t g_j^t\prod_{k\in K}\eta^t_{zjk} \Bigg]
\label{eq:resobj}\\
\text{s.t.}\quad
& \bar{y}^u_{zjk}\leq \mathcal{M}\gamma_{zjk}^u, \forall a_j\in \mathcal{A}_{dm}, k \in K, \label{eqc:uflow}\\
& \bar{y}^o_{zjk}\leq \mathcal{M}\gamma_{zjk}^o, \forall a_j\in \mathcal{A}_{dm}, k \in K, \label{eqc:oflow}\\
& \sum_{a_j\in \mathcal{A}_{dm}}\bar{y}_{zjk}^u+\bar{y}^o_{zjk}\leq \tilde{p}_{zk}, \forall k \in K, \label{eqc:ucapcity}\\
& \sum_{a_j\in \mathcal{A}_{dm}}\bar{y}^u_{zjk}\leq Q_{zk}, \forall k \in K, \label{eqc:ocapcity}\\
& (\bar{y}^u_{zjk}+\bar{y}^o_{zjk}- d_{jk})\eta^p_{zjk}=0, \forall a_j\in \mathcal{A}_{dm}, k \in K,\label{eqc:num}\\
& \bar{y}^u_{zjk}+\bar{y}^o_{zjk}\leq d_{jk}, \forall a_j\in \mathcal{A}_{dm}, k \in K,\label{eqc:num2}\\
& \bar{v}^u_{zjk} = (\tilde{\ell}_{zjk}+\tilde{o}_{zk})\gamma_{zjk}^u, \forall a_j\in \mathcal{A}_{dm}, k \in K, \label{eqc:uarrive}\\
& t_{jk}\leq \max\{\bar{v}^u_{zjk}, \beta_{zjk} \gamma^o_{zjk}\bar{v}^u_{zjk} \} +\mathcal{M}\eta^t_{zjk}, \nonumber\\
& \hspace{3.5cm} \forall a_j\in \mathcal{A}_{dm}, k \in K,\label{eqc:time}\\
& \eta_{zjk}^t, \eta_{zjk}^p, \gamma_{zjk}^u, \gamma_{zjk}^o \in \{0, 1\}, \forall a_j\in \mathcal{A}_{dm}, k \in K, \label{eqc:binary}
\end{align}
\end{subequations}

In this model, the objective in the first line of~\eqref{eq:resobj} represents the total income that the supplier can earn if its responded flows ($\bar{y}^u_{zjk}+\bar{y}^o_{zjk}$) are selected, and the risk of failing to fulfill the response, which is quantified as the product flows that exceed its production capacity (i.e., $\bar{y}^o_{zjk}$).
The second half of the objective function~\eqref{eq:resobj} indicates the reward the supplier agent will receive from the demand agents if it can satisfy the demands and deadlines.
The parameters $g_j^p$ and $g_j^t$ are the reward gains that agent $a_j$ offers if all the demands and deadlines are satisfied.
Binary variables $\eta_{zjk}^p$ and $\eta_{zjk}^t$ equal to 1 if the demand and deadline of $a_j$ for product $k$ are satisfied, 0 otherwise.
Constraints~\eqref{eqc:uflow} and~\eqref{eqc:oflow} indicate the selection of flow response.
Binary variable $\gamma_{zjk}^u$ equals 1 if $a_z$ decides to respond to $a_j$ to provide flow $\bar{y}^u_{zjk}$, and $\gamma_{zjk}^o$ equal 1 if the response from $a_z$ to $a_j$ includes a production quote that exceeds its production capacity.
Constraint~\eqref{eqc:ucapcity} indicates that the estimated production is the upper bound of the response, and constraint~\eqref{eqc:ocapcity} guarantees that $\bar{y}^u_{zjk}$ does not exceed production capacity.
Constraints~\eqref{eqc:num} and~\eqref{eqc:num2} indicate whether the response can satisfy the demand.
Equation~\eqref{eqc:uarrive} defines the arrival time based on the estimated production start time and lead time.
Constraint~\eqref{eqc:time} indicates whether the products can be delivered before the deadline.
Constraint~\eqref{eqc:binary} represents the range of all binary variables in this model.
Note that the arrival times $\bar{v}^o_{zjk}$ of over-capacity product ($\bar{y}^o_{zjk}$) cannot be smaller than the nominal arrival time (i.e., $\bar{v}^o_{zjk}= \beta_{zjk}\bar{v}^u_{zjk}\geq\bar{v}^u_{zjk}$).

\begin{figure}[tb]
\centerline{\includegraphics[width=\columnwidth]{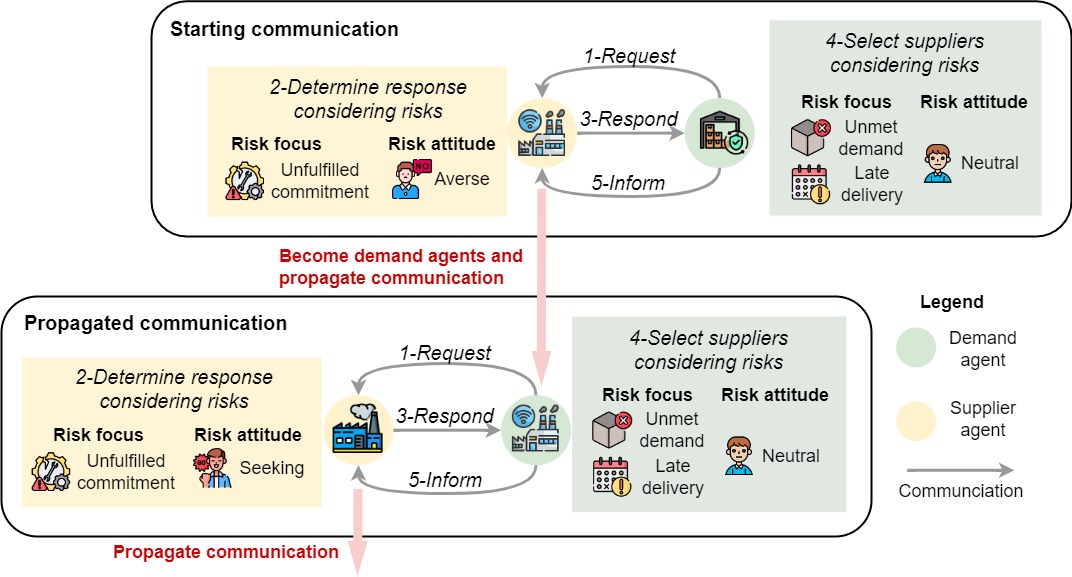}}
\caption{The agent communication and decision-making for disruption response. The agent risk focus and attitude changes for different agent roles.}
\label{fig:communication}
\end{figure}

To solve this stochastic optimization model, we apply the Sample Average Approximation (SAA) approach~\cite{kleywegt2002sample}, which generates a finite realization of the uncertain parameters following a distribution.
In this case, the known uncertain parameters include $\tilde{p}_{zk}$, $\tilde{\ell}_{zjk}$, and $\tilde{o}_{zk}$.
We denote $\xi_i=[\tilde{p}_{zk,i}, \tilde{\ell}_{zjk,i}, \tilde{o}_{zk, i}, \forall a_j\in \mathcal{A}_{dm}, k\in K]^{\mathsf{T}}$ as a vector of the sampled realizations of all the uncertain parameters which we assume to be independently distributed. Sampling $\mathcal{Q}$ realizations we define the objective as:
\begin{equation}
\vspace{-7pt}
\mathbb{E}_{1\leq i\leq\mathcal{Q}}[\mathcal{J}_s(\bar{y}_{z, i}, \bar{v}_{z, i})]=\frac{1}{\mathcal{Q}} \sum_{i}^{\mathcal{Q}}\mathcal{J}_s(\bar{y}_{z, i}, \bar{v}_{z, i})
\label{eq:estimate}
\end{equation}
and the constraints become the augmentation of all $\mathcal{Q}$ samples. 
We determine the final response $(\bar{y}_{z}^*, \bar{v}_{z}^*)$ as the expected value of the responses optimized from all the samples:
\begin{equation}
\vspace{-6pt}
\bar{y}_{z}^*=\mathbb{E}_{1\leq i\leq\mathcal{Q}}[\bar{y}_{z,i}], \ \bar{v}_{z}^*=\mathbb{E}_{1\leq i\leq\mathcal{Q}}[\bar{v}_{z,i}]
\label{eq:min}
\end{equation}
Note that objective~\eqref{eq:estimate} represents the optimization for a risk-neutral agent to minimize the expected value.
A risk-averse agent would be designed to consider the worst-case scenario, where the optimization~\eqref{eq:response} can be formulated as $\min_{1\leq i\leq\mathcal{Q}}\mathcal{J}_s(\bar{y}_z, \bar{v}_z)$.
Supplier agents solve their optimization problems independently of other supplier agents, following the distributed setting. To obtain a tractable formulation, we consider the SAA approach and solve the problems with an off-the-shelf optimization mixed-integer linear program (MILP) solver.
In the upper block in Fig.~\ref{fig:communication}, steps 2 and 3 represent the decision-making and communication process.

\subsection{Decision-making for supplier selection}
\label{sec:supplierselection}

\subsubsection{Supplier selection}
Once the demand agent $a_j$ collects the responses from the supplier agents, it solves a supplier selection optimization problem using suppliers' responses as input parameters.
The decisions include the quantity of products each supplier agent can provide, denoted by $\hat{y}_{j}=[\hat{y}_{zjk}, \forall a_z\in Z_{j}(k), k\in K]^{\mathsf{T}}$, where $\hat{y}_{zjk}$ represents the determined number of product $k$ that $a_j$ plans to get from $a_z$.
Though the response information $(\bar{y}_{z}^*, \bar{v}_{z}^*)$ is deterministic, we assume $a_j$ has different levels of trust with regards to the responses it received.
Trust is quantified as the uncertainty regarding the response that demand agent $a_j$ receives from a given supplier agent.
For example, the response from supplier agent $a_z$ includes discrete values for product quantity $\bar{y}_{z}^*$ and arrival time $\bar{v}_{z}^*$; however, the demand agent $a_j$ evaluates these variables as random variables because unexpected disturbances and variations in production and travel times exist in the real world. 
We assume that these distributions follow Gaussian distributions $\mathcal{N}(\bar{y}_{z}^*, \sigma\bar{y}_{z}^*)$, where $\sigma$ represents the trust level and is known based on prior knowledge of the agents.
Then $a_j$ evaluates the costs to receive products from each supplier along with their uncertainties about the production and delivery. 
The supplier selection optimization is formulated as a stochastic program: 

\begin{subequations}
\vspace{-9pt}
\label{eq:selection}
\begin{align}
\min_{\hat{y}_j}\ & \mathbb{E}\Bigg[C_d + w_j^t\sum_{k\in K}\Delta_{jk}^t + w_j^p\sum_{k\in K}\Delta_{jk}^p \Bigg]\label{eq:selobj}\\
\text{s.t.}\quad 
& C_d=\sum_{a_z\in Z_j(k), k \in K} m_{zjk}\hat{y}_{zjk}, \label{eqc:cost}\\
& \Delta_{jk}^t=\sum_{a_z\in Z_j(k)}\big(\lambda_{zjk}^u\max\{(\bar{v}^u_{zjk})^*-t_{jk}, 0\} \nonumber \\ & \hspace{1cm} +\lambda_{zjk}^o\max\{(\bar{v}^o_{zjk})^*-t_{jk}, 0\}\big), \forall k\in K, \label{eqc:totaltime}\\
& \Delta_{jk}^p=\max\{d_{jk}-\sum_{a_z\in Z_j(k)}\hat{y}_{zjk}, 0\}, \forall k \in K, \label{eqc:totalnum}\\
& \hat{y}_{zjk}\leq \big((\bar{y}_{zjk}^u)^*+(\bar{y}_{zjk}^o)^*\big)\lambda_{zjk}^u,  \forall a_z\in Z_j(k), k \in K, \label{eqc:finalflow}\\
& \hat{y}_{zjk}-(\bar{y}_{zjk}^u)^*\leq \mathcal{M}\lambda_{zjk}^o, \forall a_z\in Z_j(k), k \in K, \label{eqc:overflow}\\
& \lambda_{zjk}^u, \lambda_{zjk}^o \in \{0, 1\}, \forall a_z\in Z_j(k), k \in K, \label{eqc:binaryd}
\end{align}
\end{subequations}

The objective is to minimize the cost ($C_d$) to purchase the products considering the risk of unmet demand and delivery lateness due to uncertainties, as shown in~\eqref{eq:selobj}.
Equation~\eqref{eqc:cost} calculates the total cost to obtain flow $\hat{y}_{zjk}$ from the suppliers. 
Equation~\eqref{eqc:totaltime} computes the total lateness of the product delivery for all the selected suppliers, and~\eqref{eqc:totalnum} calculates the total unmet demand.
Constraint~\eqref{eqc:finalflow} indicates that agent $a_j$ cannot request more flows than what it originally supplies. 
The binary variable $\lambda^u_{zjk}$ equals 1 if supplier $a_j$ is selected for product $k$.
Constraint~\eqref{eqc:overflow} indicates whether the selected $a_j$ responds with products that exceed its production capacity.
Similarly to supply agents' optimization models, through the SAA approach, we solve a tractable MILP for demand agents with off-the-shelf state-of-the-art optimization solvers (e.g., Gurobi \cite{gurobi})
In the upper block in Fig.~\ref{fig:communication}, step 4 represents this decision-making process.
This supplier selection decision-making occurs after suppliers determine their responses.

\subsubsection{Inform selection}
Once the supplier selection decisions are made, all the demand agents $a_j$ inform each selected supplier agent $a_z$ about the new flow plan $\hat{y}_{zjk}$.
In the upper block in Fig.~\ref{fig:communication}, step 5 represents this communication process.
Note that through the described agent-based optimization, the overall problem~\eqref{eq:selection} is divided into several smaller models, which are solved based on agents' local information. 
Therefore, though the optimality of the new flow plan cannot be guaranteed in terms of the overall objective, the computational efforts are reduced. In practice, distributed solution implementation requires less coordination between agents compared to a centralized optimization model~\cite{bi2023distributed}.

\subsection{Communication propagation}
\label{sec:propagation}

Since each selected supplier agent, $a_z$, commits to providing products to meet the needs of the demand agents, this may introduce additional product/component needs from their suppliers to ensure sufficient products to meet these new commitments. 
In this case, these selected supplier agents must propagate demand requests in order to meet the needs of their related supplier agents.
The propagation process stops when all of the agents have met their additional needs (e.g., the requests have been propagated through all upstream agents). 
As shown in the lower block in Fig.~\ref{fig:communication}, these supplier agents become demand agents for further communication and decision-making.
The detailed communication propagation can be found in our prior work~\cite{bi2022model}.
In this paper, we highlight that these selected supplier agents become new demand agents with changing risk focus.

\section{Case studies}
\label{sec:casestudy}

In this paper, we focus on evaluating how agent risk attitudes affect SCN performance instead of agent interactions.
Therefore, we conduct two case studies using an out-of-sample simulation framework, which takes the new flow plans determined by agent decision-making as input.

\subsection{Set-up and simulation framework}
\label{sec:setup}

\begin{figure}[!t]
\centering
\includegraphics[width=0.9\columnwidth]{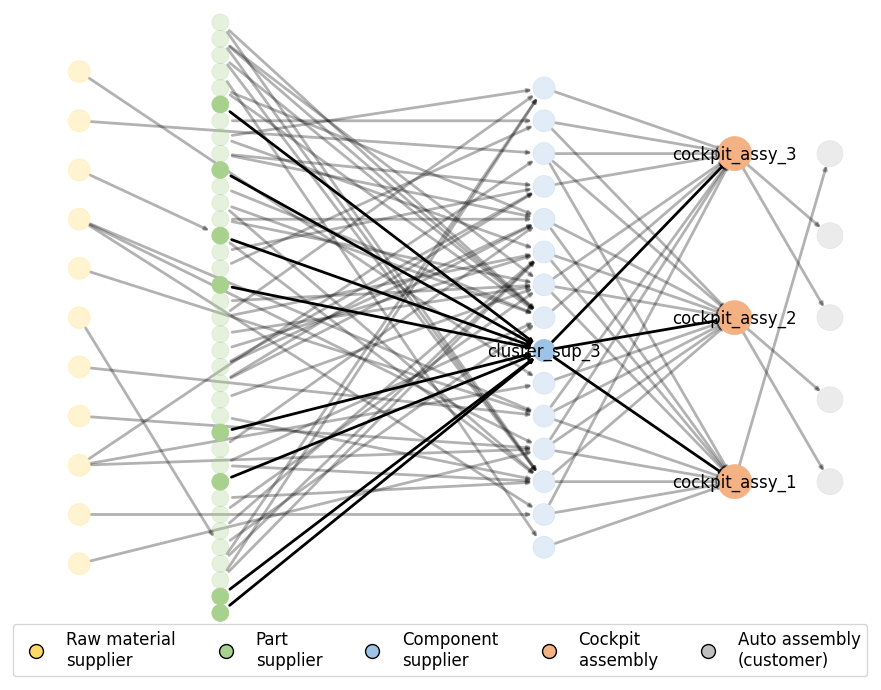}
\caption{The initial plan shown in the supply chain instance. The agents that are affected by the tested disruption are highlighted.}
\label{fig:setup}
\end{figure}

Our case studies use the cockpit supply chain instance we designed in our previous work~\cite{bi2023distributed} with additional time attributes.
In Fig.~\ref{fig:setup}, we present the topology structure of the supply chain instance with the initial flow plan.
We add lead times to all of the suppliers and cockpit assemblers, and delivery deadlines to all of the customers (i.e., auto assemblers).
We compute the initial optimal product flow plan, by solving the centralized model with the lead time in~\cite{estradagarcia2023multiobjective}, initializing the supply chain instance.

We develop an agent-based simulation framework using Python scripts for agent scheduling to evaluate disruption-response solutions under lead-time uncertainties in~\cite{estradagarcia2023multiobjective}. Given that both the realizations of the uncertainty and agent decisions follow a discrete sequence corresponding to the flow of components through the SCN, this agent-based simulation implies discrete-event dynamics.
We perform out-of-sample testing to measure the resilience of optimized response plans under the imperfect distributional knowledge available during the planning stage. The simulation is initialized with the flow plans from the upstream agents that only have outflows, starting at time 0.
The stochastic lead time for each outflow is sampled from a known normal probability distribution.
Then we obtain the delivery information, including quantity and arrival time, at all the downstream agents that just received the products. 
Note that a downstream agent may receive multiple types of products as components for its own production.
In this case, its own production starts when it receives all the needed components, i.e., the production time depends on the latest arrival time of needed components. Then the lead time for the downstream agent to deliver products to its downstream agents can be sampled.
We continue this process iteratively from upstream to downstream until the final products (i.e., cockpits) are delivered to all customers. 
The simulation runs multiple times in parallel to analyze the supply chain performance under different realizations of the lead time.

In this work, we consider a disruption that delays product delivery. 
The disrupted agent is named \textit{cluster\_sup\_3}, shown as the highlighted blue circle in Fig.~\ref{fig:setup}.
This agent provides three types of clusters, denoted by a set $K_d=\{cluster\_1, cluster\_2, cluster\_3\}$) to its downstream assembler agents, shown as the highlighted orange circles in Fig.~\ref{fig:setup}. 
For simplicity, we denote these downstream agents as $A_{dm}=\{\textit{A1, A2, A3}\}$, and \textit{A\#} represents \textit{cockpit\_sup\_\#}.
Note that there are three other cluster suppliers that could serve as backup agents.
We denote the cluster suppliers as \textit{S1, S2, S3}, and \textit{S4}, and \textit{S\#} represents \textit{cluster\_sup\_\#}.
Once a disruption is identified, the agent communication strategy is triggered to generate a new plan if necessary,  then evaluate response performance via simulation.

\subsection{Case Study 1: various disruption scales}

This case study aims to compare how the proposed approach performs when the disruption impacts the agents at different levels.
We consider three disruption scenarios, where the disruption increases the lead time of agent \textit{S3} by 20\%, 60\%, and 100\%.
In each disruption scenario, we evaluate the modified plans for instances when the three downstream agents are all risk-neutral and all risk-averse.
The out-of-sample simulation runs 300 times in parallel based on the known distributions of the lead time of all the agents.
We evaluate the performance by calculating the total delay time for when the downstream agents receive the original production flow. 
The modified local flows are represented as $[\hat{y}_{zjk}, \forall a_j \in A_{dm}, a_z\in Z_{j}(k), k\in K_d]^{\mathsf{T}}$, where $\hat{y}_{zjk}$ is the quantity of product $k$ that flows from supplier agent $a_z$ to demand agent $a_j$.
In each simulation round $i$, we denote the arrival time for flow $\hat{y}_{zjk}$ as $v_{zjk, i}$, thus the lateness of the flow is $\Delta_{zjk, i} = \max\{v_{zjk, i}-t_{jk}, 0\}$, where $t_{jk}$ is the required time for $a_j$ to receive product $k$.
The notation $\Omega_i$ represents the set of possible values of $\Delta_{zjk, i}$.
We evaluate the performance of the plan by showing the percentage of the products that have the lateness $\Delta_{zjk, i}$ for each simulation round.
The product percentage is calculated as the ratio of the total quantities in the flows that are delayed by $\Delta_{zjk, i}$ to the total product quantities:
\begin{equation}
\vspace{-5pt}
\frac{\sum_{a_j \in A_{dm}, a_z\in Z_{j}(k), k\in K_d}\hat{y}_{zjk}\ \text{if}\ \Delta_{zjk, i}=\delta_i}{\sum_{a_j \in A_{dm}, a_z\in Z_{j}(k), k\in K_d}\hat{y}_{zjk}},\ \forall \delta_i\in\Omega_i
\label{eq:lateness}
\end{equation}

\begin{figure}[!t]
\centering
\includegraphics[width=1\columnwidth]{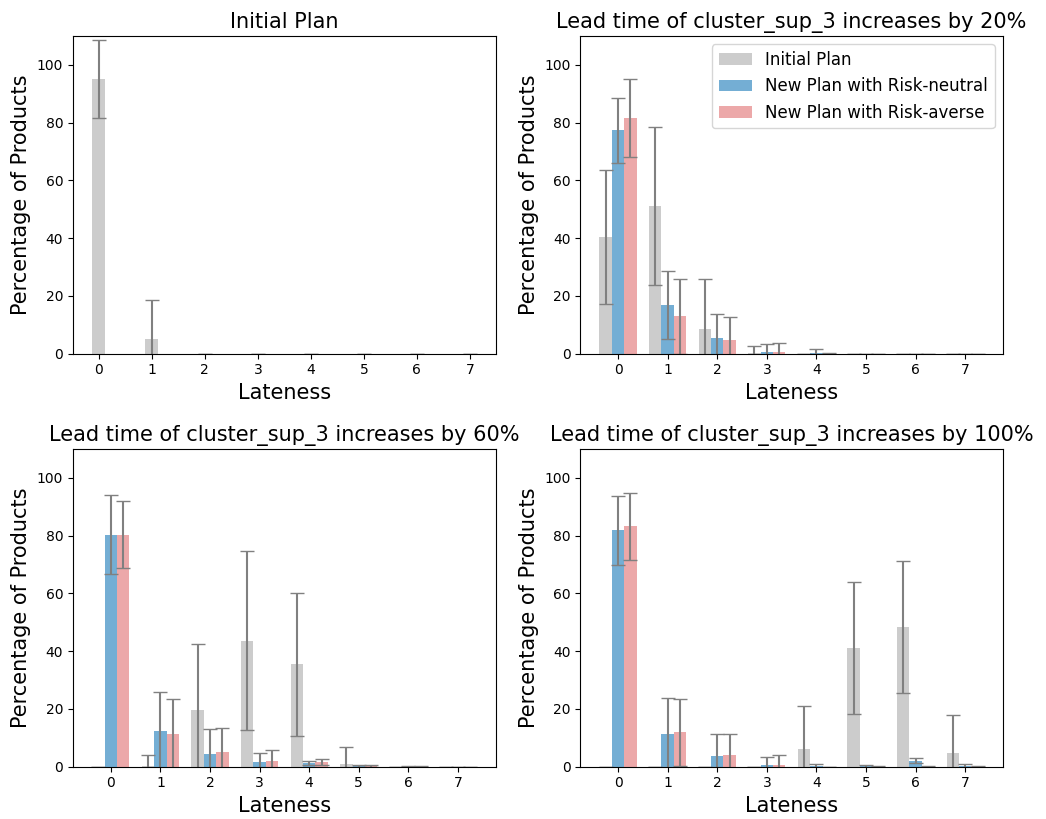}
\caption{The distribution of the total product lateness of all three downstream agents under different disruption and decision-making scenarios}
\label{fig:case1}
\end{figure}

Therefore, we can get a distribution of these percentages in terms of 300 simulation rounds.
The results shown in Fig.~\ref{fig:case1} indicate that when the disruption increases lead time more, the initial plan leads to a modified plan with more products subject to larger delays. 
When the proposed re-planning approach is applied, most of the products can be delivered on time or with a small delay.
These results demonstrate the potential reduction in delays and overall costs associated with disturbances through the application of a re-planning framework. As expected, the impact of re-planning is more pronounced for larger disruptions. 

\begin{table}
\centering
\caption{Objective values for different disruption and attitudes}
\label{tab:chpt4result}
\begin{tabularx}{\columnwidth}{cc|ccc}
\hline
\multirow{3}{*}{\textbf{Disruption}} & \textbf{Risk} & \textbf{Production} & \textbf{Lateness} & \textbf{Objective } \\
&  \multirow{2}{*}{\textbf{Attitude}}   &  \multirow{2}{*}{\textbf{Cost} $C_{dm}$}     & \multirow{2}{*}{$L_{dm}$}    & \textbf{Value} $\mathcal{J}_{dm}=$ \\
&    &      &  & $C_{dm}+w^t L_{dm}$ \\
 \hline
 
\textbf{0\%}    & \textbf{/}     & 31,120    & 0    & 31,120    \\

\multirow{2}{*}{\textbf{20\%}}   & \textbf{Neutral}    & 28,386     & 1        & 128,386   \\

 & \textbf{Averse}    & 28,416   & 1         & 128,416    \\
 
\multirow{2}{*}{\textbf{60\%}}  & \textbf{Neutral}    & 28,386    & 6         & 628,386   \\

 & \textbf{Averse}   & 29,342     & 5       & 529,342   \\
 
\multirow{2}{*}{\textbf{100\%}}  & \textbf{Neutral}   & 28,792  
& 8  & 828,792    \\

  & \textbf{Averse}   & 29,208   & 3        & 329,208\\ 
  \hline           
\end{tabularx}
\end{table}

To investigate the cost of re-planning, we check the objective values of these demand agents that are affected by the disruption.
As defined in~\eqref{eq:selobj}, each demand agent $a_j$ minimizes $\mathcal{J}_d=C_d + w_j^t\sum_{k\in K_d}\Delta_{jk}^t + w_j^p\sum_{k\in K_d}\Delta_{jk}^p$.
Note that in this case study, all the demands are satisfied, i.e., $\sum_{k\in K_d}\Delta_{jk}^p=0$, thus we focus on the product cost $C_d$ and lateness $\sum_{k\in K_d}\Delta_{jk}^t$.
We calculate the total product cost and lateness for the three demand agents:
\begin{equation}
\vspace{-5pt}
C_{dm} = \sum_{a_j\in \mathcal{A}_{dm}} C_{d, j},\ L_{dm} = \sum_{a_j\in \mathcal{A}_{dm}} \sum_{k\in K_d}\Delta_{jk}^t
\label{eq:result}
\end{equation}
In addition, the penalty weight for lateness is $w^t=10^5$ for all the demand agents, thus the total objective can be calculated by $\mathcal{J}_{dm}=C_{dm}+w^t L_{dm}$.
Table~\ref{tab:chpt4result} shows the objective values of the initial plan and modified plans for the three disruption scenarios.
As mentioned in Section~\ref{sec:setup}, in the initial plan, all the demand agents obtain clusters from \textit{S3}.
This selection is based on the performance of the entire SCN, while we focus on the cost and lateness within this subset of agents.
Note that Table~\ref{tab:chpt4result} presents the costs and penalties of the deterministic modified plan.
Fig.~\ref{fig:case1} shows the results of the simulation where the plan runs with uncertainties.

Compared to the initial plan, the modified plans have lower product costs in all scenarios.
This is because \textit{S3} has the lowest lead time and the initial plan tends to minimize the lateness due to the high penalty for lateness, even though \textit{S3} has higher product cost.
However, after the disruption occurs, \textit{S3} cannot guarantee on-time delivery.
Therefore, Table~\ref{tab:chpt4result} shows that during the re-planning process, these demand agents re-evaluate suppliers, selecting suppliers with lower product costs since lateness is inevitable.

As the disruption scale increase, the lateness increases, which indicates that the demand agents may still obtain products from the disrupted \textit{S3}.
However, when the disruption increases the lead time by 100\%, the lateness becomes smaller when the agents are risk-averse.
In this case, agents try to minimize the worst case (i.e., potential largest lead time), which mostly occurs if agents choose supplier \textit{S3}.
Therefore, agents decide to obtain products from other suppliers, even with the expense of higher production costs.
On the other hand, when agents are risk-neutral, they consider the expected value of multiple samples, thus they may still select supplier \textit{S3}, resulting in larger delays.
When the disruption scale is smaller, the uncertainties may not lead to a specific worst-case.
Consequently, the results from risk-neutral and risk-averse agents could be similar. 
Note that alternative optimization results may be achieved based on the selection of the applied weighting factors to the cost and delay penalty.

\subsection{Case Study 2: heterogeneous risk attitudes}
We conduct tests involving various combinations of risk attitudes for the three demand agents at this disruption scale to study supplier selection differences.
As mentioned in Section~\ref{sec:requestresponse}, the demand agents treat the response as normally distributed random variables, where the response is the mean value and the trust level affects the variation.

\begin{figure}[!t]
\centering
\includegraphics[width=\columnwidth]{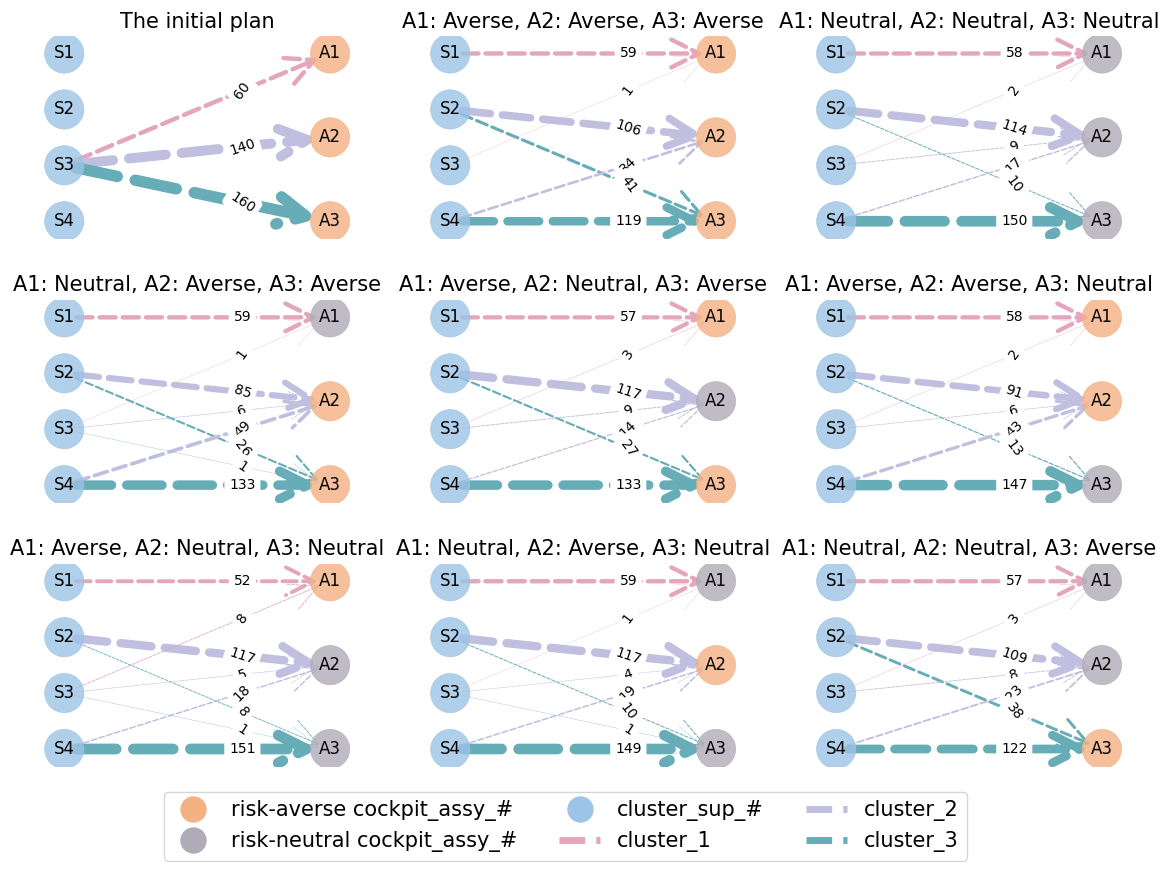}
\caption{The new flow plan for cockpit assemblers to receive clusters when they have different risk attitudes. The disruption increases 60\% of the lead time for \textit{cluster\_sup\_3} (i.e., S3 in the figure).}
\label{fig:chpt4case2}
\end{figure}

The results are shown in Fig.~\ref{fig:chpt4case2}, where the widths of the flow arrows are proportional to the quantities of products in the flow, which are labeled near the arrows.
Note that in the initial plan, all the flows to the cockpit assemblers are from \textit{S3}.
The results show that the demand agents choose to obtain products mainly from other suppliers than \textit{S3}, no matter what risk attitudes of the agents are. 
This validates the results in Fig.~\ref{fig:case1}, where most products are slightly delayed.
In general, risk-averse agents obtained fewer products from \textit{S3}.

For product \textit{cluster\_1}, agent \textit{A1} decides to switch its main supplier from \textit{S3} to \textit{S1}, regardless of its risk attitude. 
This decision is driven by several factors.
Firstly, the disruption has resulted in an increase in \textit{S3}'s lead time, making it less favorable in terms of timely product delivery. 
Secondly, \textit{S1} offers a lower cost compared to \textit{S3}. 
Lastly, \textit{A1} has a higher level of trust in \textit{S1}, meaning that there is a lower level of uncertainty associated with sourcing from \textit{S1}. 
Considering these factors, \textit{S1} emerges as the preferred choice for agent \textit{A1}, regardless of its risk attitude.

For product \textit{cluster\_2}, agent \textit{A2} decides to switch to sourcing from both \textit{S2} and \textit{S4}, with a higher volume of products from \textit{S2}.
This is because the nominal lead times of both \textit{S2} and \textit{S4} fulfill the time requirement, but \textit{S2} can produce \textit{cluster\_2} with a lower cost.
However, the lead time of \textit{S2} is larger than \textit{S4}, which introduces a higher possibility of delay.
Therefore, when \textit{A2} is risk-averse, it chooses to increase the number of products obtained from \textit{S4}.

For product \textit{cluster\_3}, agent \textit{A3} decides to switch to sourcing from both \textit{S2} and \textit{S4}, with a higher volume of products from \textit{S4}.
In this case, \textit{S4} has both a lower cost and a lower lead time.
Therefore, \textit{S4} is a preferred supplier, especially when \textit{A3} is risk-neutral.
Further, \textit{A3} holds a lower level of trust of \textit{S4}, thus it considers higher uncertainty about \textit{S4}'s response.
Therefore, when \textit{A3} is risk-averse and considers the worst case, it chooses to increase the number of products obtained from \textit{S2}.

\subsection{Discussion and insights}
\label{sec:insights}

The case studies showcased the feasibility and performance of the proposed risk management mechanism, specifically demonstrating how different risk attitudes affect agent decision-making.
This mechanism allows agents to choose different risk-based models to solve their local problems, based on their own knowledge, shared information, risk attitudes, and goals. In addition, agents can update their risk focuses and attitudes dynamically as their own attributes and local environment change.
Such flexibility in allowing agents to evaluate risk heterogeneously is difficult to achieve in a centralized approach.

The case study simulates product delivery lead-time disruptions, which can be caused by component delay or modified transportation routes. 
In addition, the simulation samples the uncertain lead time from a known distribution, while in practice, the distributional characterization of disruptions might be unknown. Given these assumptions and finite sampled scenarios, the determined new plan may not be the global optimal solution, but is optimal with respect to the sample-based reformulation.
However, the proposed framework
can be extended from different perspectives to conduct a more comprehensive investigation of heterogeneous risk management.
Firstly, disruptions on different agents can be tested to investigate how the attributes of the disrupted agents affect the decision-making when agents have different risk attitudes.
Secondly, while this work focuses on testing different risk attitudes for demand agents, introducing different risk attitudes for supplier agents can contribute to greater heterogeneity in the SCN. 
Additionally, other types of uncertainties, risks, and objectives can be considered to further investigate how agents tradeoff risks and rewards.

\section{conclusion}
\label{sec:conclusion}

Focusing on the supply chain re-planning problem, we provide a distributed decision-making approach that supports dynamic and heterogeneous risk management using a multi-agent approach.
More specifically, we reformulate the individual agent decision-making to stochastic optimization problems by incorporating uncertainties and modeling the risks that agents are interested in.
We conduct validation case studies to showcase that in a stochastic setting, the risk attitudes of agents affect their decision-making heterogeneously. 
The proposed work can be used to model the supply chain and provide decision support to handle risk heterogeneously. 
Future work will include employing multi-objective optimization techniques, extending the framework to react to simultaneous disruptions to multiple agents, and developing a hybrid strategy combining centralized approaches


\end{document}